\documentclass[submission,copyright]{eptcs}
\providecommand{\event}{ICE 2015}
\usepackage[latin1]{inputenc}
\usepackage[T1]{fontenc}

\usepackage{mathtools,amssymb,stmaryrd,amsthm,amsmath,bm}
\usepackage{color}
\usepackage{listings}
\usepackage{graphicx}
\usepackage{tabularx}
\usepackage{placeins}
\usepackage{wrapfig}
\usepackage{bigstrut}
\usepackage{booktabs}
\usepackage{lmodern}
\usepackage{subcaption}
\usepackage[yyyymmdd,hhmmss]{datetime} % for running heads

\newcommand{\grmeq}{\; ::= \;}
\newcommand{\grmor}{\; | \;}

\newcommand{\keyword}[1]{\textsf{\upshape\small #1}}
\newcommand{\protocolk}{\keyword{protocol}}

\newcommand{\valk}{\keyword{val}}

\newcommand{\messagek}{\keyword{message}}
\newcommand{\broadcastk}{\keyword{broadcast}}
\newcommand{\intk}{\keyword{integer}}
\newcommand{\floatk}{\keyword{float}}

\newcommand{\maxk}{\keyword{max}}
\newcommand{\lengthk}{\keyword{length}}
\newcommand{\truek}{\keyword{true}}

\newcommand{\foreachk}{\keyword{foreach}}
\newcommand{\ink}{\colon}

\newcommand{\gatherk}{\keyword{gather}}
\newcommand{\allgatherk}{\keyword{allgather}}
\newcommand{\reducek}{\keyword{reduce}}
\newcommand{\allreducek}{\keyword{allreduce}}
\newcommand{\scatterk}{\keyword{scatter}}
\newcommand{\andk}{\keyword{and}}
\newcommand{\mink}{\keyword{min}}
\newcommand{\sumk}{\keyword{sum}}

\newcommand{\ifk}{\keyword{if}}

\newcommand{\elsek}{\keyword{else}}

\usepackage[scaled]{beramono}

\definecolor{darkviolet}{rgb}{0.5,0,0.4}
\definecolor{darkgreen}{rgb}{0,0.4,0.2} 
\definecolor{darkblue}{rgb}{0.1,0.1,0.9}
\definecolor{darkgrey}{rgb}{0.5,0.5,0.5}
\definecolor{lightblue}{rgb}{0.4,0.4,1}

\lstdefinestyle{eclipse}{
  breaklines=true,
  basicstyle=\sffamily,
  emphstyle=\color{red}\bfseries, 
  keywordstyle=\color{black}\bfseries,
  commentstyle=\color{darkgreen},
  stringstyle=\color{darkblue},
  numberstyle=\color{darkgrey},%\lstfontfamily,
  emphstyle=\color{red},
  morecomment=[s][\color{lightblue}]{/**}{*/},
  showstringspaces=false,
  numbers=left,
  tabsize=2
}

\usepackage[scaled]{beramono}

\lstdefinelanguage{protocol}{
  style=eclipse,
  extendedchars=true,
  showstringspaces=false,
  basicstyle=\sffamily,
  morekeywords={val,size,message,send,recv,broadcast,loop,if,foreach,gather,allgather,reduce,allreduce,scatter,imessage,wait,protocol,skip},
  morekeywords={and,or,not},
  morekeywords={float,integer,positive,natural},
  morekeywords={max,min,sum,prod,nand,land,band,lor,bor,lxor,bxoer,minloc,maxloc},  % MPI_Op
   morekeywords={@synthesis, @in, @out, @condition, @exec}, %annotatios
  sensitive=false,
  morecomment=[l]{//}, 
  morecomment=[s]{/*}{*/}, 
  morestring=[b]",
  tabsize=2
}

\lstdefinelanguage{VCC}{
  language=C,
  style=eclipse,
  basicstyle=\sfffamily,
  extendedchars=true,
  showstringspaces=false,
  morekeywords={lambda,integer,result,ghost,writes,requires,ensures,pure,axiom,forall,out,assert,assume,thread_local_array,true},
  tabsize=2,
}

\lstdefinelanguage{Xtend}{
language=java,
basicstyle=\sffamily, morekeywords={cached,case,default,extension,false,import,JAVA,WORKFLOWSLOT,let,new,null,private,create,switch,this,true,reexport,around,if,then,else,context, def, dispatch, String, var, IF, ENDIF, END, ELSE},
 keywordstyle=[2]{\textbf},
 morecomment=[l]{//}, 
 morecomment=[s]{/*}{*/}, 
 morestring=[b]",
 tabsize=4}

\lstdefinelanguage{scribble}{
	style=eclipse,
	extendedchars=true,
	showstringspaces=false,
	basicstyle=\sffamily,
	morekeywords={package, type, from, as, int, global, protocol, role, choice, at, or, continue}
    sensitive=true,
    morecomment=[l]{//}, 
    morecomment=[s]{/*}{*/}, 
    morestring=[b]",
	morestring=[b]{<}{>},
    tabsize=2
}

\lstdefinelanguage{why}{
  language=[Objective]{Caml},
  style=eclipse,
  basicstyle=\sffamily,
  sensitive=true,
  showstringspaces=false,
  morekeywords={writes,theory,forall,constant,use,axiom,import,variant,invariant,requires,ensures,int},
  tabsize=2,
  morecomment=[s]{(*}{*)}
}

\title{Deductive Verification of Parallel Programs Using Why3}
\author{
    César Santos \qquad
    Francisco Martins \qquad
    Vasco Thudichum Vasconcelos
    \institute{LaSIGE, Faculty of Sciences, University of Lisbon, Portugal}
}
\def\titlerunning{Deductive Verification of Parallel Programs Using Why3}
\def\authorrunning{C.\ Santos, F.\ Martins \& V.T.\ Vasconcelos}

\begin{document}
\maketitle

\begin{abstract}
  The Message Passing Interface specification (MPI) defines a portable
  message-passing API used to program parallel computers.  MPI
  programs manifest a number of challenges on what concerns
  correctness: sent and expected values in communications may not
  match, resulting in incorrect computations possibly leading to
  crashes; and programs may deadlock resulting in wasted resources.
  Existing tools are not completely satisfactory:
  model-checking does not scale with the number of processes; testing
  techniques wastes resources and are highly dependent on the quality
  of the test set.

  As an alternative, we present a prototype for a type-based approach 
  to programming and verifying MPI-like programs against protocols. 
  Protocols are written in a dependent type language designed so as to 
  capture the most common primitives in MPI, incorporating, in addition, 
  a form of primitive recursion and collective choice. Protocols are then 
  translated into Why3, a deductive software verification tool. 
  Source code, in turn, is written in WhyML, the language of the Why3 platform, 
  and checked against the protocol. Programs that pass verification are 
  guaranteed to be communication safe and free from deadlocks.
  We verified several parallel programs from textbooks using our
  approach, and report on the outcome.
\end{abstract}

%%% Local Variables:
%%% mode: latex
%%% TeX-master: "main"
%%% End:

\section{Introduction}
\label{sec:introduction}

\lstset{language=C,style=eclipse,basicstyle=\footnotesize\ttfamily}

\paragraph{Background}

Message Passing Interface (MPI)~\cite{mpi3}, a standardized and
portable message-passing API, is the \emph{de facto} standard for
High Performance Computing. Some of the challenges in developing
correct MPI programs include: mismatches on exchanged values resulting
in incorrect computations, and deadlocks resulting in wasted time and
resources.

High performance computing bugs are quite costly. High-end HPC centers
cost hundreds of millions to commission. On many of these centers, over 3 million dollars
are spent in electricity costs alone each year and research teams
apply for computer time through competitive proposals, spending years
planning experiments~\cite{cacm}. A deadlocked
program represents an exorbitant monetary cost, and such
situations are hard to detect at runtime without resource wasting
monitors. 
%One must also consider the societal costs, since there is a
%reliance on the results of these experiments (weather simulations for
%example).

The formal verification of MPI programs employs different methodologies
such as runtime verification~\cite{must2,isp,pnmpi,dampi} and model
checking or symbolic
execution~\cite{cacm,must2,isp,siegel-vmcai11,tass-vmcai2012}. Runtime
verification, by its own nature, cannot guarantee the absence of
faults. In addition, the process can become quite expensive due to the
difficulty in producing meaningful tests. Model checking approaches
typically face a scalability problem, since the verification state
space grows exponentially with the number of processes. Verifying
real-world applications may restrict the number of processes to only a few~\cite{model-checking}.

%MPI is typically programmed in C or Fortran and is error prone,
%requiring many annotations regarding concurrency and pointer
%arithmetic if programs are to be statically
%verified.%~\cite{techreportVCC}.

\paragraph{Motivation}

To illustrate the problem we present a classic MPI example that solves
the finite differences problem.

Finite differences is a numeric method for solving differential
equations. The program starts with an initial solution $X_0$, and
calculates $X_1, X_2, X_3, \ldots$ iteratively until a maximum number
of iterations are executed. The problem vector is split amongst all
processes, each calculating their part of the problem, and then joined
at the end. The processes are setup in a ring topology as depicted
below, in order to exchange the boundary values necessary for the
calculation of the differences.

% \begin{figure}[t!][h!]
% 	\centering
\begin{center}
  \includegraphics[keepaspectratio,scale=0.9]{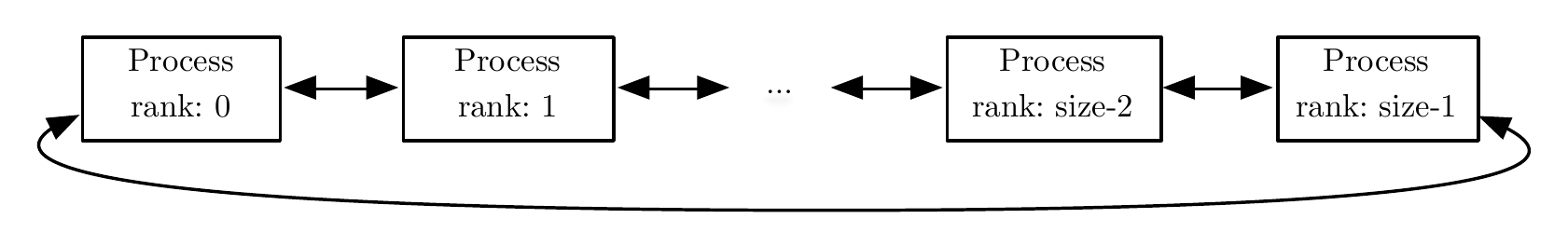}
\end{center}
% 	\caption{Processes in a ring topology}
% 	\label{fig:ring}
% \end{figure}

In MPI, every process (or participant)
is assigned a \lstinline[language=C,style=eclipse]!rank!; processes have access to their ranks. Ranks start at \lstinline[language=C,style=eclipse]!0! and end at \lstinline[language=C,style=eclipse]!size-1!, where \lstinline[language=C,style=eclipse]!size! is a constant that 
indicates the total number of processes. 
The number of processes is chosen by the user at launch time.

MPI follows the Single Program Multiple Data (SMPD) model
where all processes share the same code. Processes' behaviour diverge from
one another based on their rank (typically via conditional statements).
This model makes deployment very simple, since a single piece of code can be 
run on all machines.

Figure~\ref{fig:fdiff-c} shows an implementation of the finite
differences problem in the C programming language (simplified for
clarity). Line~2 initializes MPI, and is followed by calls to functions that return the current process rank
(line~3) and the number of processes (line~4). The size
of the input vector is broadcast (line~5), and the input vector at rank 0 is split and scattered (line~6) among all processes. Then, every processes iterates a certain number of
times, exchanging messages with its left and right neighbors (lines~9--15) and calculating local values. 
Finally, process \lstinline[language=C,style=eclipse]!0! calcutates the global error, and gathers all the parts of
the resulting vector (lines~16--17).

\begin{figure}[t!]
\begin{lstlisting}[language=VCC,style=eclipse,basicstyle=\footnotesize\ttfamily,numbers=left]
int main(int argc, char** argv) {
    MPI_Init(&argc, &argv);
    MPI_Comm_rank(MPI_COMM_WORLD, &rank);
    MPI_Comm_size(MPI_COMM_WORLD, &size);
    MPI_Bcast(&n, 1, MPI_INT, 0, MPI_COMM_WORLD);
    MPI_Scatter(data, n/size, MPI_FLOAT, &local[1], n/size, MPI_FLOAT, 0, MPI_COMM_WORLD);
    int left = rank == 0 ? size - 1 : rank - 1;
    int right = rank == size - 1 ? 0 : rank + 1;
    for (iter = 1; i <= ITERATIONS; iter++) {
        MPI_Send(&local[1], 1, MPI_FLOAT, left, 0, MPI_COMM_WORLD);
        MPI_Send(&local[n/size], 1, MPI_FLOAT, right, 0, MPI_COMM_WORLD);
        MPI_Recv(&local[n/size+1], 1, MPI_FLOAT, right, 0, MPI_COMM_WORLD, &status);  
        MPI_Recv(&local[0], 1, MPI_FLOAT, left, 0, MPI_COMM_WORLD, &status);
        // Computation is performed here, removed for simplicity
    }
    MPI_Reduce(&localErr, &globalErr, 1, MPI_FLOAT, MPI_MAX, 0, MPI_COMM_WORLD);
    MPI_Gather(&local[1], n/size, MPI_FLOAT, data, n/size, MPI_FLOAT, 0, MPI_COMM_WORLD);
    MPI_Finalize();
    return 0;
}
\end{lstlisting}
\caption{An excerpt of an incorrect implementation of the finite differences problem}
\label{fig:fdiff-c}
\end{figure}

This implementation deadlocks when using unbuffered
communication. Process \lstinline[language=C,style=eclipse]!0! attempts to send a message to its left
neighbor, which in turn is attempting to send a message to its left
neighbor, and so forth, with no process actually receiving the
message. The correct implementation of this example requires three
separate cases: one for the first process, one for the last process,
and a third for all the others. These cases have specific \lstinline[language=VCC,style=eclipse]!send!/\lstinline[language=VCC,style=eclipse]!receive!
orders that are not at all obvious.

\paragraph{Solution}

Our approach is inspired by \emph{multi-party session types}~\cite{mpst-popl,PMSTypes-fossacs}, where types describe
protocols. Types describe not only the data exchanged in messages, but
also the state transitions of the protocol and hence the allowable
patterns of message exchanges. Programs that conform to well-formed protocols
are communication-safe and free from deadlocks. Our approach makes it
possible to statically verify that participants communicate in
accordance with the protocol, thus guaranteeing the properties
above. A novel notion of type equivalence allows to type source code
for individual processes against the same (global) type.

The general idea is as follows: first, a protocol is written in a type
language designed for the purpose. A protocol compiler checks whether
the protocol is well-formed and compiles it to a format that can be
processed by a deductive program verifier.  Parallel programs are then
checked against the generated protocol.

In line with all type-based approaches, our method requires writing a ParType (a protocol)
for the program. Such a type serves as further documentation for the program. In addition, we
require a few program annotations to guide the verification tool.
%Type based approaches have the disadvantage that they require 
%not only the development of protocols (though this is not a big disadvantage as
%they also serve as documentation) but also some annotations in the program. 
%However model-checking tools typically require some annotations as well. \textbf{(citation needed, what tools, which don't...)}

\paragraph{Method}

We developed a protocol language in the form of a dependent type
language~\cite{techreport}. The language includes the most common MPI-like communication
primitives, in addition to sequential composition, primitive recursion,
and collective choice. Protocols are then translated into
Why3~\cite{why3} a deductive software verification platform that
features a rich well-defined specification language called Why.
On the other hand, source code is written in a high level language
with first class support for parallel MPI-like primitives, namely
WhyML, a language that is part of Why3.

Why3 allows the programmer to split verification conditions in parts and prove each part
using a different Satisfiability Modulo Theories (SMT) solver. In
cases where the solvers cannot handle part of the proof, Why3 can
generate code for use with proof assistants like Coq~\cite{coq}.
%
% By compiling the protocols into a format Why3 can process, programs
% can be statically verified by having all the primitives in the library
% be annotated using pre and post-conditions based on the theory of
% protocols.
%
We chose the Why3 platform in order to avoid the annotation overhead
required for static verification of C or Fortran programs, the languages
typically used to program in
MPI~\cite{techreportVCC}. This
should be considered an experiment in a new programming methodology
for developing reliable parallel applications, and not a tool for
verifying existing MPI applications.

Unlike other session type based approaches, our approach does not require explicit global-to-local protocol projection. This allows us to support not only MPMD programs, where the code for different ranks may be distinct,
but also SPMD programs such as MPI-based ones. 

\begin{figure}[t!]
\includegraphics[scale=0.9]{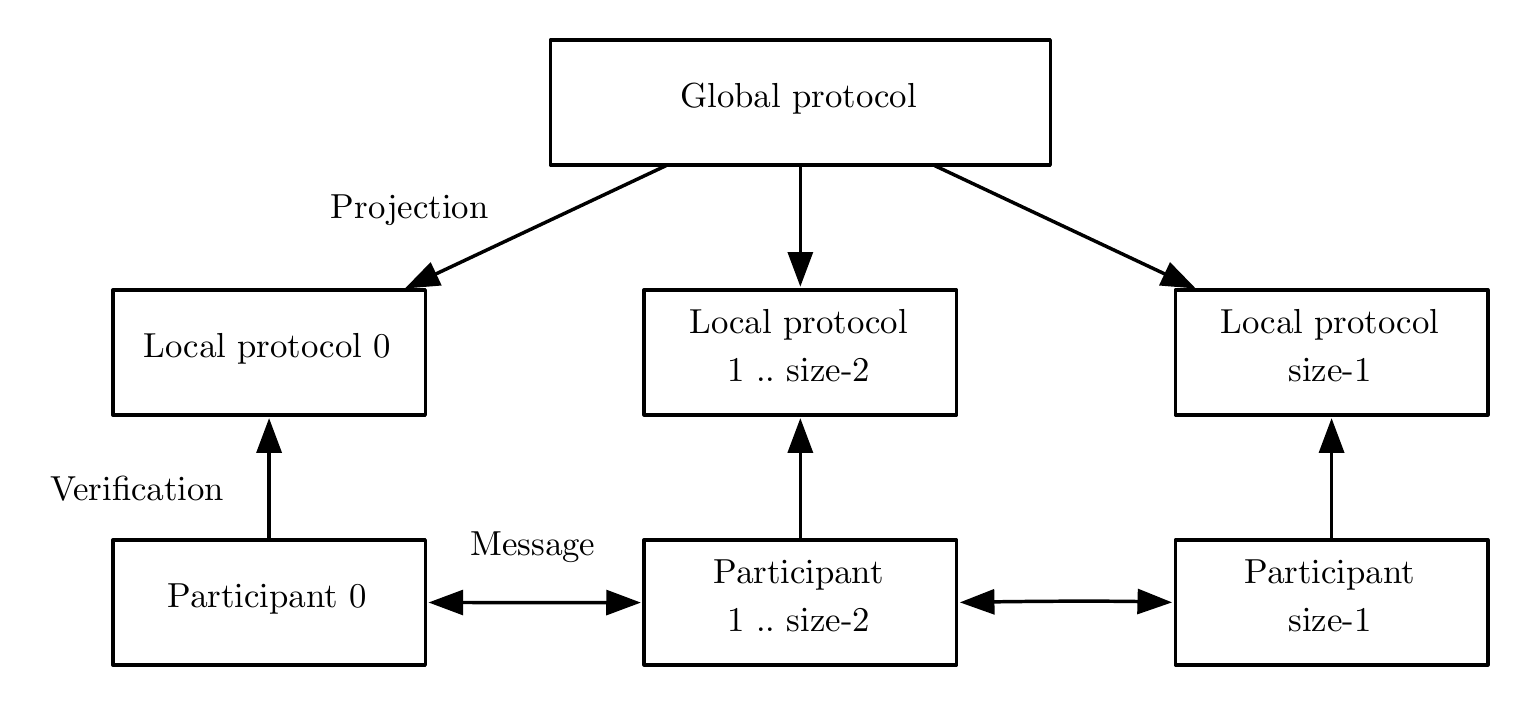}
\caption{The multi-party session types approach}
\label{fig:mpst}
\end{figure}

Figure~\ref{fig:mpst} shows the parametrized multi-party session types approach~\cite{PMSTypes-fossacs}. Global protocols are first projected into local protocols for each role, a communication pattern shared by one or more participants. In this case there are three roles, one for participant \lstinline!0!, one for participants \lstinline!1! to \lstinline!size-2!, and finally one for participant \lstinline!size-1!.

\begin{figure}[t!]
    \begin{subfigure}{0.4\textwidth}
        \includegraphics[scale=0.9]{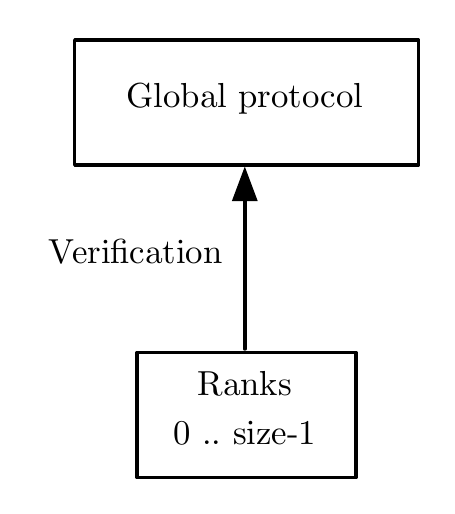}
    \caption{SPMD}
    \label{fig:ourapproachSPMD}
    \end{subfigure}%
    \begin{subfigure}{0.6\textwidth}
    \includegraphics[scale=0.9]{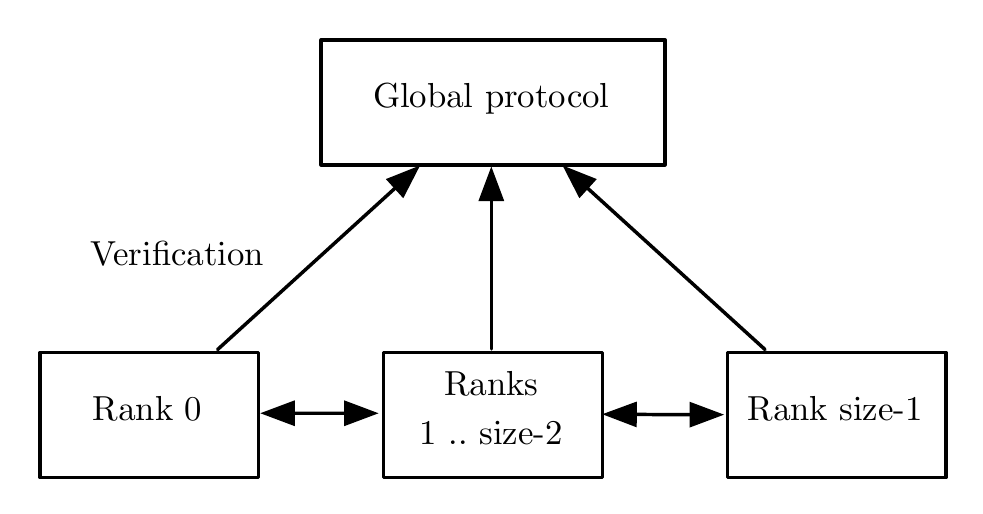}
    \caption{MPMD}
    \label{fig:ourapproachMPMD}
    \end{subfigure}
\caption{The ParTypes approach}
\label{fig:ourapproach}
\end{figure}

Figure~\ref{fig:ourapproach} shows our approach for SPMD and MPMD. Unlike multi-party session types, the ParTypes approach does not require a separate projection step: participants are verified directly against the global protocol. Furthermore, participants can be separate programs (MPMD), or single program (SPMD).

\paragraph{Contributions}

The contributions of this work are:

\begin{itemize}
\item A protocol compiler (in the form of an eclipse plugin), which
  verifies protocol formation and translates it into Why3;
\item A theory for protocols in Why3;
\item An MPI-like library for parallel programming in WhyML;
\item The verification of sample WhyML programs against protocols.
\end{itemize}

\paragraph{Outline}

The following two sections present the protocol language and our Why3
library for parallel programming, detailing the verification workflow. 
After that we present the results we obtained when comparing our approach 
against a similar tool for the C programming language. After the related work section, 
we present our conclusions and pointers to future work.

%%% Local Variables:
%%% mode: latex
%%% TeX-master: "main"
%%% End:

\section{Protocol language}
\label{sec:protocol-language}

\lstset{language=protocol,style=eclipse,basicstyle=\footnotesize\ttfamily,numbers=none}

In order to verify the finite differences example using our approach, we must first create a protocol the program must follow. The grammar for the protocol language is described in Figure~\ref{fig:grammar}. Not all protocols freely generated by the grammar are well formed. For instance, the \lstinline[language=protocol,style=eclipse]!from! and \lstinline[language=protocol,style=eclipse]!to! ranks of the \lstinline[language=protocol,style=eclipse]!message! primitive must be distinct and lie between \lstinline[language=protocol,style=eclipse]!0! and \lstinline[language=protocol,style=eclipse]!size-1!. Refer to~\cite{techreport} for details. 

\begin{figure}[t!]
 \begin{align*}
    P \grmeq& \bm{\protocolk}\; x\; p\; T & \mbox{protocol definition} \\
    T \grmeq& \bm{\messagek}\; i\; i\; D & \mbox{point-to-point comm.} \\
    \grmor & \bm{\broadcastk}\; i\; x\colon D & \mbox{broadcast operation}\\ 
    \grmor & \bm{\scatterk}\; i\; D & \mbox{scatter operation}\\
    \grmor & \bm{\gatherk}\; i\; D & \mbox{gather operation}\\
    \grmor & \bm{\reducek}\; op\; D & \mbox{reduce operation}\\
    \grmor & \bm{\allgatherk}\; x\colon D & \mbox{allgather operation}\\
    \grmor & \bm{\allreducek}\; op\; x\colon D & \mbox{allreduce operation}\\
    \grmor & \{T \; ... \; T\} & \mbox{sequence}  \\
    \grmor & \bm{\foreachk}\; x\ink i..i\; T & \mbox{repetition} \\
    \grmor & \bm{\ifk}\; p\; T\;\bm{\elsek}\; T & \mbox{collective choice} \\
    \grmor & \bm{\valk}\; x\colon D & \mbox{value} \\
    D \grmeq& \bm{\intk} \grmor \bm{\floatk} \grmor D[] \grmor \{x\colon D \mid
    p\} \grmor ... & \mbox{datatypes} \\
    i \grmeq& x \grmor n \grmor i+i \grmor \bm{\maxk}(i,i) \grmor \bm{\lengthk}(i)
    \grmor i[i] \grmor \dots & \mbox{index terms} \\
    p \grmeq& \truek \grmor i \le i \grmor p \ \bm{\andk}\ p \grmor a(i,\dots,i) \grmor  \dots & \mbox{index propositions} \\
    op \grmeq& \bm{\maxk} \grmor \bm{\mink} \grmor \bm{\sumk} \grmor \dots & \mbox{functions for reduce}
 \end{align*}
 \caption{Protocol language grammar}
 \label{fig:grammar}
\end{figure}

\begin{figure}[t!]
\begin{lstlisting}[language=protocol,style=eclipse,basicstyle=\footnotesize\ttfamily,numbers=left]
protocol FiniteDifferences (size >= 2) {
    val iterations: natural
    broadcast 0 n: {x: natural | x % size = 0}
    scatter 0 float[n]
    foreach iter: 0 .. iterations {
        foreach i: 0 .. size-1 {
            message i (size+i-1)%size float
            message i (i+1)%size float
        }
    }
    reduce 0 max float
    gather 0 float[n]
}
\end{lstlisting}
	\caption{Finite differences protocol}
	\label{fig:fdiff-protocol}
\end{figure}

The protocol for our running example is in Figure~\ref{fig:fdiff-protocol}. Every protocol specification starts with the keyword \lstinline[language=protocol,style=eclipse]!protocol! (line~1), followed by a protocol name, and a proposition (\lstinline[language=protocol,style=eclipse]!size >= 2!) that describes the number of processes required. The variable representing the number of processes is called \lstinline[language=protocol,style=eclipse]!size!, named after the MPI primitive (\lstinline[language=VCC,style=eclipse]!MPI_Size!). The protocol requires two or more processes, in order to avoid having a single process sending messages to itself, which leads to a deadlock. This restriction can be omitted and will be inferred by the validator for simple cases.

The protocol starts by specifying a global value, the maximum number of iterations. Global values are known by every process but are not exchanged by communication. They tipically represent some relevant constants hardwired in the code or present in the command line. Such values are introduced with the keyword \lstinline[language=protocol,style=eclipse]!val! (line~2). The value is given a name, \lstinline[language=protocol,style=eclipse]!iterations!, so that it may be further used.
%and a type restriction, in this case \emph{any natural divisible by the number of processes}. 
Types in the protocol language can be refined. The
language supports some abbreviations for refined types, for example \lstinline[language=protocol,style=eclipse]!natural!, which is any number greater than or equal to 0, is an abbreviation of \lstinline[language=protocol,style=eclipse]!{x: integer | x >= 0}!. The type in line~3 could instead be written 
as: \lstinline[language=protocol,style=eclipse]!{x: {y: integer | y >= 0} | x % size = 0}! or \lstinline[language=protocol,style=eclipse]!{x: integer | x >= 0 and x % size = 0}!.

All processes perform a broadcast operation, with process \lstinline!0! (the root) sending the size of the work vector to every process (line~3), followed by a \lstinline[language=protocol,style=eclipse]!scatter! operation splitting an array of size \lstinline[language=protocol,style=eclipse]{n} (line~4) among all ranks. These are examples of collective operations, where there is a root process sending data, and every process (including itself) receiving. Type \lstinline[language=protocol,style=eclipse]!float[n]! is an abbreviation of \lstinline[language=protocol,style=eclipse]!{x: float[] | length(x) = n}!. Our language allows for the specification of restrictions on the size of the array and every integer value it contains. 

The reader may have noticed that, unlike broadcast, scatter does not introduce a variable. The reason for this is that the result of a scatter operation is not identical in all processes, hence could not be referred to in the rest of the protocol.

The protocol then enters a \lstinline[language=protocol,style=eclipse]!foreach! loop (lines~5--10). The \lstinline[language=protocol,style=eclipse]!foreach! operation is not a communication primitive, but a variant of primitive recursion: a type constructor that expands its body for a given number of iterations. The inner foreach loop (lines~6--9) is used to specify the point to point communication of every process (to exchange boundary values). This constructor allows, for example, protocols to be parametric on the number of processes. Intuitively, if we were to expand the loop, it would result in the following message exchanges.

\begin{center}
\begin{tabular}{c}
\begin{lstlisting}
message 0, size-1 float         message 1, 2 float          ...
message 0, 1 float              message 2, 1 float          message size-1, size-2 float
message 1, 0 float              message 2, 3 float          message size-1, 0 float
\end{lstlisting}
\end{tabular}
\end{center}

Process \lstinline!0! first sends a \lstinline[language=protocol,style=eclipse]!message! to the process on its left, then to the process on its right (lines~7--8). Process \lstinline!1! does the same and so on until process  \lstinline[language=protocol,style=eclipse]!size-1!. Note that the protocol language does not require messages exchanges in the program to be globally sequential. Sequentiality restrictions happen on a per-process level.

The rest of the protocol is simple, process \lstinline!0! performs a \lstinline[language=protocol,style=eclipse]!reduce! operation (line~11), where every process sends process \lstinline!0! its local error, and then it calculates the global error, which is the maximum of all local errors. Finally, process \lstinline!0! collects the results of every process with a \lstinline[language=protocol,style=eclipse]!gather! operation (line~12), thus building the final solution.

The operations in the protocol are very close to those in the language, except that the language uses \lstinline[language=protocol,style=eclipse]!MPI_Send! and \lstinline[language=protocol,style=eclipse]!MPI_Recv! operations while the protocol uses \lstinline[language=protocol,style=eclipse]!message!. Protocols provide a \emph{global} point of view of the communication, while programs bear a \emph{local} point of view of the communication.

If we expand the foreach loop as a table of send and receive operations, we can easily see why the example in the introduction does not conform to the protocol: Table~\ref{tab:foreach-proj} shows the message passing pattern for every process, where we have omitted the type of the message (\lstinline[language=protocol,style=eclipse]!float!) for conciseness. There are three different orderings of \lstinline[language=protocol,style=eclipse]!send! and \lstinline[language=protocol,style=eclipse]!recv! operations: one for process \lstinline!0!, one for process  \lstinline[language=protocol,style=eclipse]!size-1!, and one for every other process. This sequencing is guaranteed not to deadlock. The example in the introduction does not match these projections.

\lstset{language=protocol,style=eclipse,basicstyle=\ttfamily}

\begin{table}
    \caption{Foreach expanded for each rank}
    \centering{\scriptsize
    \begin{tabular}{lllllll}
    \toprule
    \textbf{\lstinline!i!}  & \textbf{Loop body}
                        & \textbf{Rank \lstinline!0!}
                        & \textbf{Rank \lstinline!1!}
                        & \textbf{...}
                        & \textbf{Rank \lstinline!size-2!}
                        & \textbf{Rank \lstinline!size-1!}
                        \\
                        \midrule
    \lstinline!0!       &  \shortstack[l]{\lstinline!message 0 size-1! \\ \lstinline!message 0 1!}
                        &  \shortstack[l]{\lstinline!send size-1! \\ \lstinline!send 1!}
                        &  \lstinline!recv 0!
                        &  ~
                        &  ~
                        &  \lstinline!recv 0!
                        \\
                        \midrule
    \lstinline!1!       &  \shortstack[l]{\lstinline!message 1 0! \\ \lstinline!message 1 2!}
                        &  \lstinline!recv 1!
                        &  \shortstack[l]{\lstinline!send 0! \\ \lstinline!send 2!}
                        &  ~
                        &  ~
                        &  ~
                        \\
                        \midrule
    \lstinline!2!       &  \shortstack[l]{\lstinline!message 2 1! \\ \lstinline!message 2 3!}
                        &  ~
                        &  \lstinline!recv 2!
                        &  ~
                        &  ~
                        &  ~
                        \\
                        \midrule
    \lstinline!...!     &  ~
                        &  ~
                        &  ~
                        &  ~
                        &  ~
                        &  ~
                        \\
                        \midrule
    \lstinline!size-2!  &  \shortstack{\lstinline!message size-2 size-3! \\ \lstinline!message size-2 size-1!}
                        &  ~
                        &  ~
                        &  ~
                        &  \shortstack[l]{\lstinline!send size-3! \\ \lstinline!send size-1!}
                        &  \lstinline!recv size-1!
                        \\
                        \midrule
    \lstinline!size-1!  &  \shortstack[l]{\lstinline!message size-1 size-2! \\ \lstinline!message size-1 0!}
                        &  \lstinline!recv size-1!
                        &  ~
                        &  ~
                        &  \lstinline!recv size-1!
                        &  \shortstack[l]{\lstinline!send size-2! \\ \lstinline!send 0!}
                        \\
                        \bottomrule
    \end{tabular}}
    \label{tab:foreach-proj}
\end{table}

\lstset{language=protocol,style=eclipse,basicstyle=\footnotesize\ttfamily}
\section{Why3 theory for protocols}
\label{sec:programming-language}

\lstset{language=why,style=eclipse,basicstyle=\footnotesize\ttfamily,breaklines=false}

With the protocol out of the way, we concentrate on programming the algorithm. This ordering is not a requirement, the protocol could have been written by a different programmer, or both the protocol and the program could be developed simultaneously.

To enable the verification of MPI programs with Why3, we developed two libraries: the Why3 theory for protocols that provides a representation for protocols as a Why3 datatype and the WhyML MPI library that replicates part of the MPI API with pre and post-conditions for the various communication primitives. 

\paragraph{Why3 theory for protocols} The Why3 theory for protocols features a representation of protocols as a Why3 datatype (Figure~\ref{fig:protocoldata-why}). Every datatype constructor either has a continuation or takes the rest of the protocol as a parameter (except for \lstinline[language=why]!Skip!, the empty protocol), unlike the protocol language where primitives are sequenced with the sequencing operator ($\{T \; ... \; T\}$ in Figure~\ref{fig:grammar}). Continuations are implemented using the \lstinline[language=why]!HighOrd! theory of Why3, to allow values from the program to be introduced into the protocol during verification. Protocols in Why3 format are generated from the protocol language by a translator that first checks the good formation of types generated by the grammar in Figure~\ref{fig:grammar}. Each datatype constructor corresponds to a type constructor. Primitives that introduce values use the continuation format, while those that do not feature the rest of the protocol as a parameter. Both kinds feature a datatype ($D$ in Figure~\ref{fig:grammar}) representing the data exchanged.

The type \lstinline[language=why]!pred a! used in the datatype constructors (lines~17--20) is an abbreviation of \lstinline[language=why]!func a bool!, a function of any type to a boolean. Such a predicate is used to restrict values, encoding refinement datatypes ($\{x \colon D \mid p \}$ in Figure~\ref{fig:grammar}) This type abbreviation is part of the Why3 standard library. Similarly, the type \lstinline[language=why]!cont a! used in the continuation constructors (lines~23--26) is an abbreviation of \lstinline[language=why]!func a protocol!, a function of any type to a protocol (which is the continuation, used to introduce values into the protocol).

\begin{figure}[t!]
\begin{lstlisting}[language=why,style=eclipse,basicstyle=\footnotesize\ttfamily,numbers=left]
type protocol = 
      Val datatype continuation
    | Broadcast int datatype continuation
    | AllGather datatype continuation
    | AllReduce op datatype continuation
    | Scatter int datatype protocol
    | Gather int datatype protocol
    | Message int int datatype protocol
    | Reduce int op datatype protocol
    | If datatype protocol protocol protocol
    | Foreach int int (cont int) protocol
    | Skip
with
    op = Max | Min | Sum | Prod | Land | Band | Lor | Bor | Lxor | Bxor
with
    datatype =
      IntPred (pred int)
    | FloatPred (pred float)
    | ArrayIntPred (pred (array int))
    | ArrayFloatPred (pred (array float))
with    
    continuation =
      IntAbs (cont int)
    | FloatAbs (cont float)
    | ArrayIntAbs (cont (array int))
    | ArrayFloatAbs (cont (array float))
with 
    cont = func a protocol
\end{lstlisting}
	\caption{The Why3 datatype for protocols}
	\label{fig:protocoldata-why}
\end{figure}

%To work with the datatypes, there are two main functions: \lstinline[language=why]!matches! which checks that a value is true for the predicate in the datatype and \lstinline[language=why]!apply! which applies the value to the continuation. The implementations of these functions can be seen in Figure~\ref{fig:matches-why}. They are implemented using axioms because otherwise there would be a type error when calling the predicate or continuation. In this way the functions remain generic while still verifying correctly.
%
%\begin{figure}[t!][ht!]
%	\lstinputlisting[language=why,style=eclipse,basicstyle=\footnotesize\ttfamily,numbers=left]{examples/matches.why}
%	\caption{Matches and apply}
%	\label{fig:matches-why}
%\end{figure}

%Other utility functions are implemented, mostly to extract parts of the protocol like the body of \lstinline[language=why]!ForEach!, or to get the next primitive in a protocol.

\begin{figure}[t!]
\begin{lstlisting}[language=why,style=eclipse,basicstyle=\footnotesize\ttfamily,numbers=left]
val send (dest:int) (v:'a) (s:state) : ()
	writes { s.protocol }
	requires { 0 <= dest /\ dest < size }
    requires { match project s.protocol with 
    		   | Message src dst d _ -> src = rank /\ dest = dst /\ matches v d
    		   | _ -> false 
               end }
	ensures { s.protocol = next (old s).protocol }
    
val apply (v:'a) (s:state) : 'a
    writes { s.protocol }
    requires { match project s.protocol with 
               | Val d _ -> matches v d
               | _ -> false 
               end }
    ensures { s.protocol = continuation (old s).protocol v }
    ensures { result = v }
    
val foreach (s:state) : foreach_data
    writes { s.protocol }
    requires { match project s.protocol with 
               | Foreach _ _ _ _ -> true 
               | _ -> false 
               end }
    ensures { s.protocol = next (old s).protocol }
    ensures { result = (
        foreach_body (old s).protocol,
        foreach_from (old s).protocol,
        foreach_to (old s).protocol )}   
        
val expand (fd:foreach_data) (i:int) : state
    requires { let _,low,high = fd in low <= i <= high }
    ensures { let body,_,_ = fd in result = {protocol = body i} }
    
function project (p:protocol) : protocol =
    match p with
        | Message source dest _ remainingProtocol -> 
            if rank <> source /\ rank <> dest 
                then project remainingProtocol 
                else p
        | _ -> p
    end
\end{lstlisting}
	\caption{WhyML MPI-like library (excerpt)}
	\label{fig:primitives}
\end{figure}

\begin{figure}[t!]
\begin{lstlisting}[language=why,style=eclipse,basicstyle=\footnotesize\ttfamily,numbers=left]
let main () =
    let s = init fdiff_protocol in
    let iterations = apply 100000 s in
    let n = broadcast 0 input s in
    let local = scatter 0 work s in
    let left = if rank > 0 then rank-1 else size-1 in
    let right = if rank < size-1 then rank+1 else 0 in
    for iter = 1 to iterations do
        let inbody = expand (foreach s) iter in (* Annotation *)
        let body = foreach inbody in (* Annotation *)
        if (rank = 0) then (
            let f1 = expand body 0 in (* Annotation *)
            send left local[1] f1;
            send right local[n/size] f1; isSkip f1; (* Annotation *)
            let f2 = expand body 1 in (* Annotation *)
            local[n/size+1] <- recv right f2; isSkip f2; (* Annotation *)
            let f3 = expand body (np-1) in (* Annotation *)
            local[0] <- recv left f3; isSkip f3); (* Annotation *)
        else if (rank = size-1) then (
            let f1 = expand body 0 in (* Annotation *)
            local[lsize+1] <- recv right f1; isSkip f1; (* Annotation *)
            let f2 = expand body (np-2) in (* Annotation *)
            local[0] <- recv left f2; isSkip f2; (* Annotation *)
            let f3 = expand body (np-1) in (* Annotation *)
            send left local[1] f3;
            send right local[lsize] f3; isSkip f3); (* Annotation *)
        else (
            let f1 = expand body (rank-1) in (* Annotation *)
            local[0] <- recv left f1; isSkip f1; (* Annotation *)
            let f2 = expand body rank in (* Annotation *)
            send left local[1] f2;
            send right local[lsize] f2; isSkip f2; (* Annotation *)
            let f3 = expand body (rank+1) in (* Annotation *)
            local[lsize+1] <- recv right f3; isSkip f3); (* Annotation *)
        isSkip inbody; (* Annotation *)
        (* Computation is performed here, removed for simplicity *)
    done;
    globalerror := reduce 0 Max !localerror s;
    gather 0 local s;
    isSkip s; (* Annotation *)
\end{lstlisting}
	\caption{The WhyML program for the corrected finite differences example}
	\label{fig:fdiff-fixed}
\end{figure}

\paragraph{WhyML MPI library} The WhyML MPI library includes MPI-like primitives such as \lstinline[language=why]!init!, \lstinline[language=why]!broadcast!, \lstinline[language=why]!scatter!, \lstinline[language=why]!gather!, \lstinline[language=why]!send!, \lstinline[language=why]!recv!, as well as \emph{annotations} \lstinline[language=why]!foreach!, \lstinline[language=why]!expand!, and \lstinline[language=why]!isSkip! required to guide the verification process.
In order to check that the program follows the protocol, each Why3/MPI primitive is annotated with pre and post-conditions.

The \lstinline[language=why]!init! primitive initializes the verification state, a structure used during the verification process. The verification state has a single mutable field containing the protocol datatype. Every ParTypes primitive takes the verification state as a parameter, verifies that the protocol is correct for the primitive and, if there are no errors, updates the protocol field of the verification state with the protocol continuation. Some of the ParTypes primitives and their annotations can be seen in Figure~\ref{fig:primitives}. 

Every ParTypes primitive calls the utility function
\lstinline!project! when verifying the protocol
(lines~35--42). Obtaining the projection of a protocol yields the
protocol itself in most cases. The exception is when the protocol is a
message that is neither originating in nor addressed to the rank in
question. In this case the function recurs, skipping messages
unrelated to the current rank. Note that rank is not passed as a parameter, instead the \lstinline!rank! constant is handled automatically by the Why3 verification process.

The \lstinline[language=why]!send! primitive (lines~1--8), receives the target rank, a value, and the current state. It checks that the destination is a valid process (line~3), that the current verification state starts with a \lstinline[language=why]!Message! after calling the project function (with the current rank as the source and the same destination as in the program), and that the value being sent matches the refinement (line~5). The \lstinline[language=why]!send! primitive returns nothing and ensures that the next state is the protocol after the message (line~8).

The \lstinline[language=why]!apply! primitive is used to introduce program values into the protocol. It checks that the head of the protocol is a \lstinline[language=why]!Val!, and that the value introduced matches the restriction (line~13). The contract ensures that the protocol becomes the continuation of the \lstinline[language=why]!Val! constructor, after applying the value (line~16). Since a value is introduced, this primitive uses \lstinline!continuation! instead of \lstinline!next!, which does not introduce values.

Finally, the \lstinline[language=why]!foreach! primitive, requires that the protocol must have a \lstinline[language=why]!Foreach! constructor at the head (line~22). It ensures that the protocol continues with whatever follows \lstinline[language=why]!Foreach! (line~25), updates the verification state, and returns a triple containing the body of the foreach and its range (lines~26--29). The \lstinline[language=why]!expand! function takes the output of the \lstinline[language=why]!foreach! primitive and an integer~\lstinline[language=why]!i!, checks that the integer is in range (line~33), and returns the projection of the \lstinline[language=why]!Foreach! for that integer (line~34). 

The ParTypes primitives \lstinline!apply! and \lstinline!foreach! are not related to any MPI primitive, they are simply annotations required to match the program against a protocol.

%Normally the head of the protocol is the primitive at the start, but that is not true if the primitive at the start is a \lstinline[language=why]!Message!. Messages are only relevant for the processes they refer to, so for unrelated ranks the message exchanges in the protocol should be skipped. The head function recursively checks if the start of the protocol is a message, checks if the current rank being verified is not one of the communicating processes in the message, and removes the message from the protocol if that is the case.There are more primitives but they function similarly to these and we, therefore, omit them.

\paragraph{Checking WhyML code} Figure~\ref{fig:fdiff-fixed} shows the
finite differences example written in WhyML, following the three
separate send/receive orderings. The necessary annotations are all marked
with \lstinline[language=why]!(* Annotation *)!. On line~2, the
verification state is initialized with the finite differences
protocol, the result of translating the protocol in
Figure~\ref{fig:fdiff-protocol} into a Why3 protocol type (an instance
of the type in Figure~\ref{fig:protocoldata-why}, not shown). On
line~3, the \lstinline[language=why]!apply! primitive is used to
introduce the number of iterations into the protocol, consuming the
Val constructor at the head. The subsequent lines perform a
\lstinline[language=why]!broadcast! and a
\lstinline[language=why]!scatter! following the protocol, while
consuming the corresponding constructors.

On line~10 the \lstinline[language=why]!foreach! and \lstinline[language=why]!expand! primitives are called to obtain the body of the Foreach constructor, with the current iteration count applied. The Foreach body contains another Foreach loop (see Figure~\ref{fig:fdiff-protocol}), but that loop does not correspond to a loop in the program, instead, it is used to define the behavior of every process. There are three different \lstinline[language=why]!send!/\lstinline[language=why]!recv! orderings, following the message projections in Table~\ref{tab:foreach-proj}. Each branch (lines~11--34) corresponds to one of the projections, with the \lstinline[language=why]!Foreach! constructor expanded for all intervening processes: the process on the left, the process itself, and the process on the right (lines~12, 15 and 17 for example). The \lstinline[language=why]!isSkip! function verifies that each of these projections is equivalent to Skip at that point in the code (lines~14, 16 and 18 for example), to guarantee the protocol is followed correctly (see~\cite{techreport} for the theory). The rest of the program is simple, with a final \lstinline[language=why]!isSkip! at the end (line~40) to guarantee the protocol was completely consumed.

%%% Local Variables:
%%% mode: latex
%%% TeX-master: "main"
%%% End:

\section{Evaluation}
\label{sec:evaluation}

We adapted a few classic parallel programming examples to WhyML, wrote their protocols, and checked them with Why3. To evaluate the results, we compared verification times and the ratio of annotations or of code. The closest work to ours checks programs written in C+MPI with VCC~\cite{techreportVCC}. The experimental setup was a 2,4 GHz Intel Core 2 Duo machine running Windows~7 with 4 GB of RAM.

\paragraph{Sample programs} We verified the following programs:

\begin{itemize}
    \item \textbf{Pi}: a simple program that calculates an approximation of pi through numerical integration, taken from~\cite{using-mpi}.
    \item \textbf{Finite differences}: used as our running example. The code is adapted from~\cite{foster}.
    \item \textbf{Parallel dot}: calculates the dot product of two vectors, taken from~\cite{pacheco}.
\end{itemize}

\paragraph{Verification time} The verification times obtained are in Table~\ref{tab:results-time} (average of 10 runs). Though Why3 can be used interactively, all the runs were automated, and no manual proofs were necessary. As can be seen from the results, Why3 and VCC have similar performance. This is surprising as Why3 spawns a different Z3 process for each sub-proof. A possible explanation for the similarity is that each individual sub-proof is substantially easier on the solver, and VCC has to perform more verifications than Why3 due to concurrency and pointer related proofs.

The results are promising, more so since proofs can be done in parts if necessary.

\begin{table}
    \centering{\scriptsize
    \begin{tabular}{lrrrr}
    \toprule
    \textbf{Program}   &\textbf{Why3 Sub-Proofs}  & \textbf{Why3 Time (s)} & \textbf{VCC Time (s)} & \textbf{Why3/VCC} \\
    \midrule
    Pi                 & 27             & 1,6          & 2,4           & 66,7\%           \\
    Finite differences & 374            & 14,9          & 16,1           & 92,5\%           \\
    Parallel dot       & 298            & 7.9          & 7,4           & 106,7\%           \\
    \bottomrule
    \end{tabular}}
    \caption{Results for Why3 and VCC verification times}
    \label{tab:results-time}
\end{table}

\paragraph{Annotation effort} The ratio for annotations can be seen in Table~\ref{tab:results-anot}. The lines of code (LOC) count ignores library imports, comments, and empty lines. VCC requires more annotations than Why3 due to concurrency and pointer related annotations, but a lot of these can be automated with an annotator or by employing C macros. That said, something similar could be done for Why3.
%
%Passing the protocol state through the various MPI primitives is done explicitly in Why3 while in VCC this is avoided by using macros that hide the parameter. This could be solved in Why3 with an external annotator or using a higher level language that would then compile to WhyML.
%
The only annotations the programmer would have to write would be foreach and collective choice marks, greatly reducing the effort required to use our verification methodology.

\begin{table}
    \centering{\scriptsize
    \begin{tabular}{lrrrrrr}
    \toprule
    \textbf{Program}   & \textbf{Why3 LOC} & \textbf{Why3 Anot} & \textbf{Ratio} & \textbf{VCC LOC} & \textbf{VCC Anot} & \textbf{Why3/VCC} \\
    \midrule
    Pi                 & 33           & 6           & 18\%           & 42            & 10          & 23\%         \\
    Finite differences & 86           & 29          & 33\%           & 128            & 49         & 38\%         \\
    Parallel dot       & 61           & 11          & 18\%           & 99            & 30         & 30\%         \\
    \bottomrule
    \end{tabular}}
    \caption{Results for Why3 and VCC annotation requirements}
    \label{tab:results-anot}
\end{table}
\section{Related Work}
\label{sec:related-work}

\textbf{Scribble}~\cite{scribble} is a language to describe protocols for
message-based programs based on the theory of multiparty session
types~\cite{mpst-popl}.  Protocols written in
Scribble include explicit senders and receivers, thus ensuring that
all senders have a matching receiver and vice versa. Global protocols
are projected into each of their participants' counterparts, yielding
one local protocol for each participant present in the global
protocol.  Developers can then implement programs based on the local
protocols and using standard message-passing libraries, like
Multiparty Session C~\cite{session_c}.

Pabble~\cite{pabble-pdp} is a parametric extension of
Scribble, which adds indices to participants and represents Scribble
protocols in a compact and concise notation for parallel
programming. Pabble protocols can represent the interaction patterns
of scalable MPI programs, where the number of participants in a
protocol is decided at runtime through parameters.

In this work we depart from multiparty session types along two
distinct dimensions: a) our protocol language is specifically built
for MPI primitives, and b) we do not explicitly project a protocol but
else check the conformance of code directly against a global protocol.

\paragraph{Tools for the verification of MPI programs}
The objectives of MPI verification tools are diverse
and include the validation of arguments to MPI primitives as well as
resource usage~\cite{dampi}, ensuring interaction properties such as the
absence of deadlocks~\cite{isp,siegel-vmcai11,dampi}, or asserting
functional equivalence to sequential
programs~\cite{siegel-vmcai11}. The methodologies
employed are also diverse, ranging from traditional dynamic
analysis up to model checking and symbolic execution.  In comparison,
our methodology is based on type checking and deductive program
verification, thus avoiding testing and the state-explosion problem inherent to
the model checking approaches described below.

\paragraph{Model checking}
TASS~\cite{tass-vmcai2012} employs model checking and
symbolic execution, but is also able to verify user-specified
assertions for the interaction behavior of the program, so-called
collective assertions, and to verify functional equivalence between
MPI programs and their sequential counterparts. The
approach performs a number of checks besides deadlock detection (such
as buffer overflows and memory leaks), but, as expected, does not
scale with the number of processes.

MOPPER~\cite{mopper-fm14} is a verifier that detects deadlocks by
checking formulae satisfiability, obtained by analyzing execution
traces of MPI programs. It uses a propositional encoding of
constraints and partial order reduction techniques, obtaining
significant speedups when compared with ISP.  The concept of parallel
control-flow graphs~\cite{dfa-ics13} allows for the static and dynamic
analysis of MPI programs, e.g., as a means to verify sender-receiver
matching in MPI source code.

CIVL~\cite{siegel-etal:2014:civl-tr} is a model checker that uses a
C-like unified intermediate verification language for specifying
concurrency in message-passing, multi-threaded, or GPU languages. The
tool uses model checking techniques and symbolic execution to detect
deadlocks, assertions and bounds violations, as well as illegal memory
usages. The tool's underlying CVC3 theorem prover may fail due to
state space explosion, and the user needs to specify input bounds on
the command line to specify a finite subset of state space.

\paragraph{Runtime verification and testing}
ISP~\cite{isp} is a deadlock detection tool that explores all possible
process interleaving using a fixed test harness. 
Dynamic execution analyzers, such as DAMPI~\cite{dampi} and
MUST~\cite{must2}, strive for the runtime detection of deadlocks and
resource leaks.

\section{Conclusion}
\label{sec:conclusion}

We developed an eclipse plugin for the development and validation of protocols, and a programming language for the development of parallel programs by adding MPI-like primitives to WhyML. We also developed a Why3 theory of protocols for the verification of MPI-like WhyML programs. With this approach we can ensure that programs that pass the Why3 verification are free from deadlocks, all message exchanges are type safe, and the program adheres to the protocol.

Unlike model checkers (such as TASS~\cite{tass-vmcai2012}), our approach scales to any number of processes, running in constant time. No runtime verification of the software is necessary as in ISP~\cite{isp}, DAMPI~\cite{dampi} or MUST~\cite{must2}. These tools do not require protocols and typically require less program annotations, but the runtime verifiers require a good test suite which is much harder to write than a protocol.
Unlike Scribble~\cite{scribble}, our approach can model MPI-like programs, including collective choices without communication.

Previous work used VCC~\cite{vcc} to verify C+MPI programs. The approach is very similar to ours, but requires extra annotations regarding concurrency and pointers since VCC is a tool for verifying concurrent C programs. The annotations are also more complex, while in our approach are more natural and fit with the code. 

The \lstinline[language=why]!foreach! primitive annotation requires familiarity with how the \lstinline[language=why]!foreach! primitive is expanded, but writing correct programs already implies that sort of mental reasoning. Na\"{i}ve approaches most likely result in deadlocks, as we illustrated in the finite differences example (Figure~\ref{fig:fdiff-c}). The VCC based approach shares these problems.

We successfully verified a number of textbook examples of parallel programs, with verifications taking only a few seconds in the worst case, and none of the examples required manual proofs. 

Our prototype is based on a verification language, WhyML, that is not an appropriate language for industry use. While OCaml programs can be extracted from WhyML, OCaml is not a language typically used in high performance computing. Performance is the major consideration in high performance computing, and Fortran and C are the fastest high-level languages available. To tackle these issues (including the annotation problem), an appropriate language should be developed. This language, like our WhyML based language, would have first class parallel programming primitives, but it would be essentially a C or Fortran superset with restrictions. This language would compile to either C or Fortran, and by having essentially the same semantics, performance should be the same.

%A major hurdle would be existing legacy code, so at least some level of interoperability with existing C or Fortran code is a must. One way to make this language appealing to HPC developers, besides the guarantee of deadlock freedom, would be the possibility of optimizing MPI primitives into faster, but unsafer, alternatives while guaranteeing the program still behaves correctly.

Finally, other MPI primitives need to be supported, such as asynchronous communication primitives, topologies, communicators and wild card receive.

\begin{sloppypar}
  \paragraph{Acknowledgments.} This work is supported by FCT through
  project Advanced Type Systems for Multicore Programming and project
  Liveness, Statically (PTDC/EIA-CCO/122547 and 117513/2010) and the
  LaSIGE lab (PEst-OE/EEI/UI0408/2011). We would like to thank
  Dimitris Mostrous for his insightful comments.
\end{sloppypar}

\bibliographystyle{eptcs}

%\clearpage
%\appendix

%\section{Listings}
%\label{programs}

%\subsection{Pi}
%\label{app:fdiff-computation-header}
%\lstinputlisting[language=why,style=eclipse,basicstyle=\footnotesize\ttfamily,numbers=left]{examples/pi-full.mlw}

%\subsection{Finite differences}
%\label{fdiff-full}
%\lstinputlisting[language=why,style=eclipse,basicstyle=\footnotesize\ttfamily,numbers=left]{examples/fdiff-full.mlw}

%\subsection{Parallel dot}
%\label{app:fdiff-computation-impl}
%\lstinputlisting[language=why,style=eclipse,basicstyle=\footnotesize\ttfamily,numbers=left]{examples/paralleldot-full.mlw}

%\section{Protocols}
%\label{protocols}

%\subsection{Pi}
%\lstinputlisting[language=protocol, style=eclipse,basicstyle=\footnotesize\ttfamily,numbers=left]{examples/pi.prot}

%\subsection{Finite differences}
%\lstinputlisting[language=protocol, style=eclipse,basicstyle=\footnotesize\ttfamily,numbers=left]{examples/fdiff.prot}

%\subsection{Parallel dot}
%\lstinputlisting[language=protocol, style=eclipse,basicstyle=\footnotesize\ttfamily,numbers=left]{examples/paralleldot.prot}

\end{document}

%%% Local Variables:
%%% mode: latex
%%% TeX-master: t
%%% End: